\documentclass[aps,pra,showpacs,superscriptaddress,preprint]{revtex4-1}
\usepackage{graphicx}
\usepackage{amsmath,bm}

\begin{document}

\title{Basic mechanisms in the laser control of non-Markovian dynamics.}

\author{R. Puthumpally-Joseph}
\affiliation{Laboratoire Interdisciplinaire Carnot de Bourgogne (ICB), UMR 6303 CNRS-Universit\'e Bourgogne Franche Comt\'e, 9 Av. A. Savary, BP 47 870, F-21078 Dijon cedex, France}
\affiliation{Institut des Sciences Mol\'eculaires d'Orsay (ISMO) UMR CNRS 8214 ,Universit\'e Paris Saclay, Univ. Paris Sud,  F-91405 Orsay, France}

\author{E. Mangaud}
\affiliation{Laboratoire Collisions Agr\'egats R\'eactivi\'e (IRSAMC), Universi\'e Toulouse III Paul Sabatier, UMR 5589, F-31062 Toulouse Cedex 09, France}

\author{V. Chevet}
\affiliation{Laboratorie Chimie Physique (LCP)-CNRS,Universit\'e Paris Saclay, Univ. Paris Sud,  F-91405 Orsay, France}

\author{M. Desouter-Lecomte}
\affiliation{Laboratorie Chimie Physique (LCP)-CNRS,Universit\'e Paris Saclay, Univ. Paris Sud,  F-91405 Orsay, France}
\affiliation{D\'epartement de Chimie, Universit\'e de Li\`ege, Sart Tilman, B6, B-4000 Li\`ege, Belgium}

\author{D. Sugny}
\affiliation{Laboratoire Interdisciplinaire Carnot de Bourgogne (ICB), UMR 6303 CNRS-Universit\'e Bourgogne Franche Comt\'e, 9 Av. A. Savary, BP 47 870, F-21078 Dijon cedex, France}
\affiliation{Institute for Advanced Study, Technische Universit\"at M\"unchen, Lichtenbergstrasse 2 a, D-85748 Garching, Germany}

\author{O. Atabek}
\affiliation{Institut des Sciences Mol\'eculaires d'Orsay (ISMO) UMR CNRS 8214, Universit\'e Paris Saclay, Univ. Paris Sud,  F-91405 Orsay, France}

\date{\today}
\begin{abstract}
Referring to a Fano-type model qualitative analogy we develop a comprehensive basic mechanism for the laser control of the non-Markovian bath response in strongly coupled Open Quantum Systems (OQS). A converged Hierarchy Equations Of Motion (HEOM) is worked out to numerically solve the master equation of a spin-boson Hamiltonian to reach the reduced electronic density matrix of a heterojunction in the presence of strong THz laser pulses. Robust and efficient control is achieved increasing by a factor $\times 2$ non-Markovianity measured by the time evolution of the volume of accessible states. The consequences of such fields on the central system populations and coherence are examined, putting the emphasis on the relation between the increase of non-Markovianity and the slowing down of decoherence processes.

\end{abstract}
\maketitle

\section{Introduction}
Non-unitary dynamics among quantum states of a sub-system coupled to its environment involving a large number of degrees of freedom, where dissipation and decoherence evolve simultaneously is the basic concern of the theory of Open Quantum Systems (OQS)~\cite{Weiss,Breuer,BreuerLPV,Rivas,Vega,Krauss, Alicki}. Since no physical system can truly be considered as isolated, OQS are actually very common not only in physics, but also in chemistry and biology, where they have recently attracted considerable attention in applications ranging from quantum technologies in condensed phase, to electronic and proton transfers in flexible proteins~\cite{MayKuhn}. But even more important is to build quantum control strategies to optimize physical observables \cite{Glaser,koch2} such as decoherence rates or efficient and fast charge transfers over large molecular structures, with final challenges as crucial as light harvesting in photosynthetic organisms~\cite{Chin2013,Gelinas2014,Scholes2011}.

When aiming at control OQS over a wide range of time, energy or temperature, the environmental bath response to the central system can no longer be neglected. As a consequence, this induces large memory effects with non-Markovian evolution describing dissipation and decoherence that should appropriately be taken into account~\cite{BreuerLPV,Rivas}. In other words, any control exerted on the central system would be limited in time and robustness by the unavoidable dissipation towards the bath~\cite{Glaser,koch2,Potz,Potz2,Lidar,Cui,Tai}. The ultimate challenge should be to take advantage of the back-flow of information characterizing non-Markovianity, to enforce the control of the central system physical observables~\cite{Koch, Poggi}. Said differently, the question of at what extent appropriately controlling memory effects (i.e., non-Markovianity or entropy, for instance) fighting against decoherence, would affect the robustness of the central system characteristics protecting them from dissipation, is of major interest. Several steps towards this goal are in order: (i) Acting on the central system only, through a strong static dc field producing a Stark shift among the eigenenergies of the two-level system, enhance non-Markovianity is expected from off-resonant excitation as has been shown in \cite{molphys}. The present work is precisely dedicated to the generalization of such strategies to intense laser field controls. Moreover, we are observing the consequences of increased non-Markovianity on the central system populations and coherence; (ii) Still acting on the central system only, but now using an optimal control scheme aiming at some protection against decoherence of its physical characteristics (population revival or robust exchange), observe the consequences in terms of the bath non-Markovian response. This analysis which goes beyond the scope of this paper will be published elsewhere~\cite{NJPoct2017}; (iii) Locally control in an appropriate way the increase of some non-Markovianity witnesses (volume of accessible states, entropy or free-energy) and relate this response with its expected consequences, such as less dissipative central system observables. This future prospect, i.e., taking advantage from non-Markovianity for a robust control of the full system dynamics, would presumably require for its efficiency, acting on both the central system and its environmental bath in a direct way. This could be reached by referring to some collective modes which guide the flow of information from the central system to the bath in a reversible manner~\cite{BreuerLPV}. It is however to be noted that even if the control field is assumed to explicitly have a dipole interaction with the central system only, it still has an influence on the bath dynamics through a memory kernel involved in the master equation driving the central system evolution \cite{Meier,Ohtsuki,Yan}.
The paper is organized as follows. In Sec.~\ref{theory}, a spin-boson Hamiltonian \cite{Weiss,Leggett} is worked out referring to realistic parameters taken from a model heterojunction between fullerene and oligothiophene molecules \cite{Tamura,Tamura2,Mangaud}. The dynamics of the central system density matrix described by a non-Markovian master equation is solved using the so-called hierarchical equations of motion (HEOM) up to convergence~\cite{Kubo,Ishizaki,Tanimura}. Stark shift as a basic mechanism for non-Markovianity control is introduced through a qualitative Fano-type analogy. The time-evolution of the volume of accessible states illustrated on an appropriate Bloch sphere is taken as a measure of non-Markovianity~\cite{lorenzo2013}. Section~\ref{results} is dedicated to the presentation of the results. The external control fields are chosen both as realistic ultra short duration dc flashes, or THz single optical cycle pulses with intensities less than $5 \times 10^{12} W/ cm^2$. The major result is a spectacular enhancement of non-Markovianity that could be achieved through a tunable Stark shift, which thus turns out to be a basic control mechanism for such OQS. Finally, the response on such a control of the central system physical observables (populations and coherence) is investigated. Additionally, Supplemental Materials are provided online at \cite{SM} to illustrate the time evolution of the volume in field-free and field-controlled cases. 

\section{Model and Theory}\label{theory}

\subsection{The spin-boson Hamiltonian}
Having in mind donor-acceptor type of charge transfer processes, we consider a fullerene-oligothiophene heterojunction modeled by a molecular dimer within a two-level approximation making up the central system $S$~\cite{Tamura,Tamura2,Mangaud}. More precisely, a spin-like Hamiltonian $H_S$ describes two electronic states $|1\rangle$ and $|2\rangle$ of a diabatic representation, radiatively coupled through a dipole interaction. The 2$\times$2 matrix representation of $H_S$ (in a.u.) is given by:
\begin{equation}
H_S(t) = \delta \sigma_z+W\sigma_x-\bm{\mu} E(t)
\label{HS}
\end{equation}
$\sigma_x, \sigma_z$ are the corresponding Pauli matrices, $2\delta$ measures the diabatic energy gap between $|1\rangle$ and $|2\rangle$, with $W$ the interstate potential electronic coupling. Their actual values are those corresponding to the heterojunction with an interfragment distance fixed at $R=2.5 \AA$, leading to $2\delta =0.517\, {\rm eV} $ and $W= 0.2\, {\rm eV}$. This amounts to an eigenenergy gap of $\omega_0 = 0.654\, {\rm eV}$ (that is $0.024$a.u.) in the adiabatic basis obtained by diagonalization $W$, with a corresponding Rabi period of 6.3 fs. As for the dipole matrix $\bm{\mu}$, it is the only quantity entering the model that is not yet calculated from quantum chemistry codes. For our donor-acceptor system, we model it, in the diabatic basis, as a diagonal matrix
\begin{equation}
\bm{\mu}  =\mu_0 \sigma_z = \mu_0\,\begin{bmatrix}
1 & 0 \\
0 & -1
\end{bmatrix}
\end{equation}
assuming a value $\mu_0=1$ a.u.
Finally, the time-dependent electric field amplitude is noted $E(t)$ and the resulting time-dependence of $H_S(t)$ occurs only through the radiative coupling in length gauge, $-\bm{\mu} E(t)$, with the additional assumption $\bm\mu$ (as a vector) be aligned with the linearly polarized electric field~\cite{seideman2003}. It is worthwhile noting that, even if the experimental feasibility of such an alignment is questionable, the ultra-short pulse durations we are referring to are such that the molecular fragment rotational dynamics can safely be assumed as frozen.

All nuclear degrees of freedom involved in the vibronic description of the heterojunction are associated with a bosonic bath. More precisely, this bath collects all normal modes of the two oligothiophene-fullerene fragments. The bosonic time-independent part of the Hamiltonian is written in terms of (mass-weighted) nuclear coordinates $q_k$ and their associated momenta $p_k$, $k$ labeling a given harmonic oscillator associated to a normal mode:
\begin{equation}
H_{boson} = \frac{1}{2} \sum_{k=1}^{N}[p_k^2+\omega_k^2(q_k \pm d_k/2)^2]
\label{Hboson}
\end{equation}
$d_k$ are the spatial shifts between equilibrium geometries in the two electronic states. Actually, $d_k$'s are responsible for the central system-bath couplings (vibronic couplings) as is clear when displaying $H_{boson}$ in three terms:
\begin{equation}
H_{boson} = H_B+H_{SB}+H_{ren}
\label{threeterms}
\end{equation}
with
\begin{equation}
H_B=\frac{1}{2} \sum_{k=1}^{N}[p_k^2+\omega_k^2 q_k^2],
\label{HB}
\end{equation}
\begin{equation}
H_{SB}=B \sigma_z,  \,     B=\sum_{k=1}^{N} c_kq_k,
\label{HSB}
\end{equation}
and
\begin{equation}
H_{ren}= \sum_{k=1}^{N} (c_k/\sqrt{2} \omega_k)^2.
\label{Hren}
\end{equation}
$H_B$ is the bath Hamiltonian, $H_{SB}$ is the system-bath coupling $B$ being a collective bath coordinate with vibronic coupling coefficients $c_k= \omega_k^2 d_k/2$ involving $d_k$ and $H_{ren}$ is an energy renormalization. In the following, $N=264$ normal modes are retained and their frequencies $\omega_k$ are assumed to be the same in both electronic states $|1\rangle$ and $|2\rangle$. It is shown that an alternate way to fully characterize the spin-boson coupling is through a spectral density written as a frequency comb \cite{Breuer}:
\begin{equation}
\mathcal{J}(\omega) = \frac {\pi}{2} \sum_k \frac{c_k^2}{\omega_k}\delta(\omega-\omega_k)
\label{comb}
\end{equation}
In our heterojunction case, the spectral density is given as a continuous functional form ~\cite{Meier,Chenel2}
\begin{equation}
\mathcal{J}(\omega) = \sum_{k=1}^{M} \dfrac{\omega p_{k}}{\left[(\omega-\Omega_k)^2 + \Gamma_k^2\right]\left[(\omega+\Omega_k)^2 + \Gamma_k^2\right]}\,,
\label{Eq:jw}
\end{equation}
with all fit parameters (up to $M=5$) provided in Table \ref{Table_fit}.

\begin{table}[ht]
	\caption{parameters for spectral density $ \mathcal{J}(\omega) $.}
	\begin{ruledtabular}
		\begin{tabular}{c c c }
			     $p_k$ (a.u)              &   $\Omega_k$ (a.u)         & $\Gamma_k$ (a.u)       \\ [0.5ex]
			\hline \noalign{\vskip 1.0ex}
			 $ 3.72\times 10^{-10}  $     &   $ 6.99\times 10^{-3} $  & $ 5.86\times 10^{4} $ \\ [0.5ex]
			 $ 1.90 \times 10^{-11} $     &   $ 3.05\times 10^{-3} $  & $ 5.50\times 10^{4} $ \\ [0.5ex]
			 $ 7.80\times 10^{-12}  $     &   $ 4.00\times 10^{-3} $  & $ 4.70\times 10^{4} $ \\ [0.5ex]
			 $ 5.80 \times10^{-12}  $     &   $ 1.94\times 10^{-3} $  & $ 6.83\times 10^{4} $ \\ [0.5ex]
			 $ 8.00\times10^{-12}   $     &   $ 5.20\times 10^{-3} $  & $ 7.00\times 10^{4} $ \\ [0.5ex]
			\hline \noalign{\vskip 1.0ex}
		\end{tabular}
	\end{ruledtabular}
	\label{Table_fit}
\end{table}

\noindent
In summary, apart from the dipole matrix, all parameters entering the spin (energy gap and residual diabatic interstate coupling) and bosonic parts (spectral density) are those of the heterojunction characterized by its interfragment geometry.

\subsection{The non-Markovian master equation}

The key observable in OQS dissipative dynamics is the reduced density matrix $\rho$ which is given as the partial trace, over bath degrees of freedom, of the full density matrix $\Xi$:
\begin{equation}
\rho(t)=Tr_B[\Xi(t)]
\label{trace}
\end{equation}
Projection techniques used within Nakajima-Zwanzig~\cite{Breuer} formalism lead to a non-Markovian master equation which could be recast as:
\begin{equation}
\partial_t\rho(t)=\mathcal{L}_{eff}(t)\rho(t)+\int_0^tdt'K(t,t')\rho(t')
\label{master}
\end{equation}
where the effective Liouvillian reads:
\begin{equation}
\mathcal{L}_{eff}(t)\rho(t) =-i+[(H_s(t)+H_{ren}), \rho(t)]
\label{liouvillian}
\end{equation}
and $K(t,t')$ is the already mentioned memory kernel.
The solution of Eq.~(\ref{master}) requires an initial condition for which a separability between the central system and the bath is assumed at $t=0$:
\begin{equation}
\Xi(0)=\rho(0)\rho_{eq},
\label{separability}
\end{equation}
$\rho_{eq}$ being the bath density matrix at thermal equilibrium.
For complex systems the most challenging part is the numerical evaluation of the memory kernel. In this work we are using a well documented strategy based on hierarchy equations of motion (HEOM)~\cite{Kubo,Ishizaki,Tanimura}. One of the requirements of this method, based on path integral techniques, is an exponential expansion of the correlation function of the collective bath mode $B$ defined in Eq.~(\ref{HSB}). Actually, the fluctuation-dissipation theorem relates the correlation function $\mathcal{C}$ to the bath spectral density $\mathcal{J}$~\cite{Breuer}:
\begin{equation}
\mathcal{C}(t,t_0)=\frac{1}{\pi}\int_{-\infty}^{+\infty}\frac{e^{-i\omega(t-t-0)}}{1-e^{-\beta\omega}}\mathcal{J}(\omega)d\omega
\label{CJ}
\end{equation}
where the bath temperature $T$ enters in $\beta=1/k_BT$, $k_B$ being the Boltzmann factor. With the two-pole Lorentzian form of $\mathcal{J}$, referring to Cauchy's residues theorem when evaluating the integral in Eq.~(\ref{CJ}), the correlation function and its complex conjugate are finally written as \cite{Pomyalov}
\begin{equation}
\mathcal{C}(t)=\sum_k\alpha_ke^{i\zeta_kt},
\label{C}
\end{equation}
and
\begin{equation}
\mathcal{C}^*(t)=\sum_k\tilde{\alpha}_ke^{i\zeta_kt}.
\label{C*}
\end{equation}
The solution of Eq.(\ref{master}) turns out to be the first element of a chain of auxiliary density matrices $\rho_n(t)$ obeying a system of coupled equations written as \cite{}:
\begin{eqnarray}
{\dot \rho _{\bf{n}}}(t) & = & - i\left[ {{H_S}(t),{\rho _{\bf{n}}}(t)} \right] + i\sum_k {{n_k}{\zeta _k}{\rho _{\bf{n}}}\left( t \right)} \nonumber \\
& - & i\left[ {\sigma_z,\sum_k {{\rho _{{\bf{n}}_k^ + }}\left( t \right)} } \right] \nonumber \\
& - & i\sum_k {{n_k}\left( {{\alpha _k}\sigma_z{\rho _{{\bf{n}}_k^ - }} - {{\tilde \alpha }_k}{\rho _{{\bf{n}}_k^ - }}\sigma_z} \right)}
\label{Eq:HEOM}
\end{eqnarray}
${\bf{n}_k} = (n_1,...n_K)$ being a vector giving the occupation numbers in the $K$ dissipative modes involved in the decomposition of $\mathcal{C}^*(t)$ and  $ {\bf{n}}_k^ {\pm}  = \left\{ {{n_1}, \cdots ,{n_k} \pm 1, \ldots ,{n_{{n_{cor}}}}} \right\} $. The level $L$ of an auxiliary matrix in the hierarchy corresponds to the sum $L=\sum_{k=1}^K n_k$. The mathematical structure is such that each density matrix of level $L$ is coupled to matrices of level $L\pm1$ and $L=0$ leads to $\rho(t)=\rho_{\bf{0}}$ with all occupation numbers zero. Equation (\ref{Eq:HEOM}) is solved at a given level of hierarchy $L$, corresponding to a given approximation.
It is worthwhile noting that the first moment of the collective mode $B$, given as $X^{1}=Tr_B[B\Xi(t)]$, may help for a better understanding of the correlated system-bath dynamics. The HEOM formalism provides a direct evaluation of $X^{1}$ in terms of the first level auxiliary matrices (with $\bf{n}$ such that $\sum_k{n_k} = 1$) \cite{Mangaud}:
\begin{equation}
X^{1}(t)=-\sum_{\bf{n}} \rho_{\bf{n}}(t)
\label{X1}
\end{equation}
In particular, the importance of memory effects in Eq.(\ref{master}) can directly be probed through this first moment upon recasting it in the master equation, leading finally to \cite{Shi}:
\begin{equation}
\partial_t\rho(t)=-i[H_S(t),\rho(t)] +i[\sigma_z, X^{1}(t)]
\label{X1r}
\end{equation}

\begin{figure}[ht]
	\includegraphics[width =0.6\columnwidth]{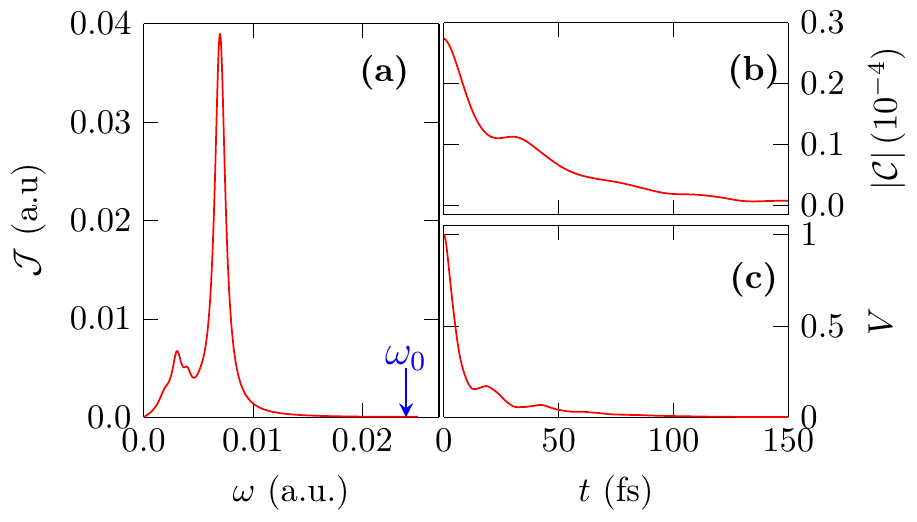}
	\caption{(Color online) Panel (a) shows the spectral density $ \mathcal{J}(\omega) $. Panel (b) is the corresponding correlation function $ \mathcal{C} (t)$ at $ T =300 $ K and (c) is the volume of accessible states $ V(t) $. The gap between the two levels $\omega_0 = 0.024au$ is indicated by the blue vertical arrow in (a)}
	\label{fig:system-bath}
\end{figure}

\subsection{Basic mechanism for control}
The spectral density $\mathcal{ J}(\omega)$ is displayed in Fig.~\ref{fig:system-bath}a and its corresponding correlation function $\mathcal{C}(t)$ at room temperature in Fig.~\ref{fig:system-bath}b. In the absence of an external field, the two adiabatic eigenlevels of the central system are only indirectly coupled through their environmental bath. $\mathcal{ J}(\omega)$ can be viewed as a frequency representation of an energy dependent discrete-continuum coupling scheme appropriately averaged over the density of levels of the discretized quasi- continuum. At that respect, the spin-boson model could be put in analogy with a standard Fano resonance representation of two discrete states (central system) facing and interacting \cite{Fano, Finkelstein} (through $\mathcal{ J}(\omega)$) with a discretized quasi-continuum (set of bath harmonic oscillators). A highly structured $\mathcal{ J}(\omega)$, as the one of Fig.~\ref{fig:system-bath} with two well-peaked Lorentzians, is expected to lead to important memory effects. Such narrow peaks could be attributed to some long-lived Feshbach resonances that are locally modifying the density of states. They are supporting bath collective modes, presumably enhancing their moments, and due to their long enough lifetimes could temporarily trap the system-bath dynamics, or efficiently mediate it, ultimately leading to enhanced non-Markovianity. Even more important for the control purpose, is the central system transition frequency $\omega_0$ which can be progressively tuned through the variation of the external field amplitude $E(t)$. As has previously been discussed, two cases are in consideration: on-resonant, when $\omega_0$ matches one of the two maxima of $\mathcal{ J}(\omega)$, off-resonant otherwise~\cite{clos2012}. The latter is expected to produce the most important memory effects. This can be rationalized once again referring to a Fano model analogy. Actually for the off-resonant case the central system is only weakly coupled to the bath (low values of the spectral density corresponding to $\omega_0$). The back flow of information from the bath to the central system is organized along two strategies in competition: (i) either on-the-energy-shell, i.e., at the same frequency that $\omega_0$, but with a low system-bath coupling; (ii) or off-the-energy-shell, i.e., at frequencies corresponding to local maxima of the spectral density, with large system-bath couplings. The last strategy requires absorption of additional photons or excitation of phonons and proceeds from longer times leading to non-Markovianity. The specific peaked structure of the spectral density offers thus a control flexibility by tuning $\omega_0$ from on- to off- resonant cases. As has been previously suggested, this is expected to be achieved through an external field producing a fully controlled Stark shift among the levels of the central system via their transition dipole~\cite{molphys}. The strategy followed hereafter is precisely based on this adaptable Stark shift, taken as a basic mechanism. However, the control field indirectly affects the bath dynamics and does not completely disentangle the action over the central system from the bath.

\subsection{Non-Markovian evolution of the volume}

Several witnesses of non-Markovianity have recently been discussed in the literature~\cite{BreuerLPV,BreuerLP,clos2012}, among which is the volume of accessible states~\cite{Paternostro}, the associated decoherence rate of a time-dependent Lindblad-type evolution~\cite{Anderson,Anderson2} and the von Neumannentropy \cite{Haseli}. More precisely, the time evolution of the central system density matrix can be written through a quantum dynamical map $F(t)$ and an initial condition:
\begin{equation}
\rho(t)=F(t)\rho(0)  ,\\\rm{for}   \\\ t\ge 0
\label{map1}
\end{equation}
Assuming non-singularity of $F(t)$ at time $t$, differentiating Eq.~(\ref{map1}) one gets a time-local matrix equation:
\begin{equation}
\partial_t\rho(t)=\mathcal{L}(t)\rho(t)=\dot{F}F^{-1}\rho(t)
\label{map2}
\end{equation}
The non-Markovian character is associated with the relaxation rate of the generator $\mathcal{L}$. The volume of accessible states $V(t)$ is obtained by mapping the density matrix on its corresponding Bloch sphere using the complete orthogonal basis set of Pauli matrices together with the identity. The time evolution of the volume of this Bloch ball, $V(t)$, is then given by the determinant of $F(t)$, a quantity independent of any initial condition:
\begin{equation}
V(t)=\textrm{det} [F(t)].
\label{volume}
\end{equation}
It can also be shown that the total decoherence rate $\Gamma(t)$ of $\mathcal{L}$ is related to the volume through~\cite{Anderson}:
\begin{equation}
V(t)=V(0) \exp [-2\int_0^t \Gamma't')dt'].
\label{gamma}
\end{equation}
Finally, the dynamics is said to be non-Markovian if:
\begin{equation}
\frac{dV(t)}{dt} \ge 0
\label{criterion}
\end{equation}
or equivalently, if $\Gamma(t) < 0$, as opposite to a situation where the total decay rate is constant. A back-flow of information from the bath to the central system can be observed for values of $\Gamma(t)$, temporarily negative, for which the time evolution of the volume departs from pure exponential decay and may even show some bumps which turn out to be clear signatures of non-Markovianity ~\cite{lorenzo2013}.

\section{Results and discussion}\label{results}

The numerical results are presented in three sub-sections discussing the following aspects: (i) Determination of generic field parameters and convergence of the associated dynamical evolution calculations with respect to successive orders of HEOM; (ii) Evolution of the volume and of the total decoherence rate as resulting from specific control fields; (iii) Consequences of the control of non-Markovianity on the time evolution of physical observables such as the populations of the initial state, the coherence and the bath collective modes.

\subsection{HEOM convergence}

\begin{figure}[ht]
	\centering
	\includegraphics[width =0.6\columnwidth]{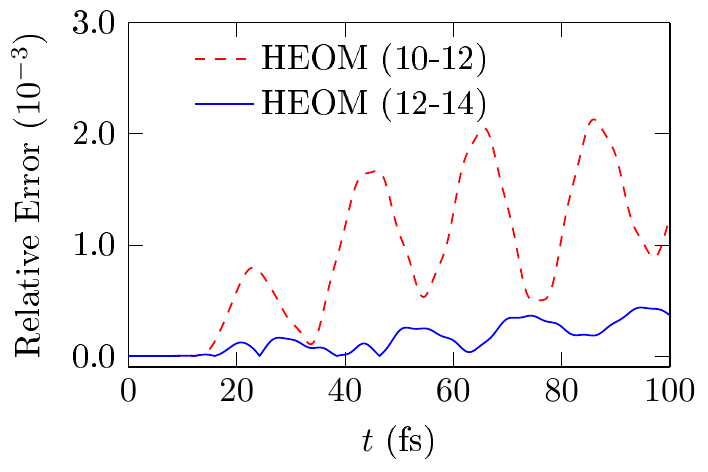}
	\caption{(Color online) Relative error in $ V(t) $ calculated for successive $2L$ orders of HEOM. Red dashed curve is the error in $ V(t) $ between order $ 10 $ and $ 12 $, whilst the blue solid curve is the same for order $ 12 $ and $ 14 $}
	\label{fig:convergence}
\end{figure}

Fig.~\ref{fig:system-bath} illustrates together with the spectral density and the corresponding evolution of the correlation function, the time-dependent volume of accessible states, in field-free conditions. A typical time scale for the overall decay process could be estimated as about $60$~fs after which the correlation function has decayed to almost one third of its initial value and the volume to almost zero.
Our first purpose is to check the numerical convergence when solving Eq.~(\ref{Eq:HEOM}) as a function of increasing level $L$ of hierarchy which corresponds to perturbation order $2L$. This is done, in field-free situation, evaluating the relative error affecting the volume, that is $[V((2L+2)-V(2L)]/V(2L)$, when increasing $2L$. The results are displayed in Fig.~\ref{fig:convergence} as a function of time up to $100$~fs.
For the perturbation order $2L=10$ the relative error remains less than $2 \times 10^{-3}$ for the overall dynamics with some oscillations occurring at about $25$~fs and $45$~fs  roughly corresponding to times leading to a plateau behavior of the correlation function. When increasing the level of hierarchy, at perturbation order $2L=12$, the relative error is clearly attenuated and does no more exceed $0.5 \times 10^{-3}$ which seems to be acceptable for an overall characterization of the volume. To avoid highly time-consuming calculations we fix $2L=12$ for the convergence criterion. The required perturbation order in HEOM basically depends on the importance of the central system-bath coupling, given by the spectral density. The inclusion of a control field, even strong, within the central system would only indirectly affect these couplings. The relevance of the convergence criterion set for the field-free case has successfully been checked for the field driven dynamics.

Finally, it is also interesting to note that the volume is not decaying monotonously but shows two bumps of modest amplitude at about $25$~fs and $40$~fs. These are signatures of non-Markovianity presumably due to an off-resonant configuration and specific spectral density of our model heterojunction. Actually, the transition frequency $\omega_0$ is larger than the frequency corresponding to the maximum amplitude of  $\mathcal{J}(\omega)$, $\omega_{max}=0.007$~a.u. Our control goal is to enhance the non-Markovian signature (amplitude of the bumps) by Stark shifting the energy levels of the central system, tuning  $\omega_0$ through the application of a control field.

\subsection{Control fields}

\begin{figure}[ht]
	\centering
	\includegraphics[width =0.6\columnwidth]{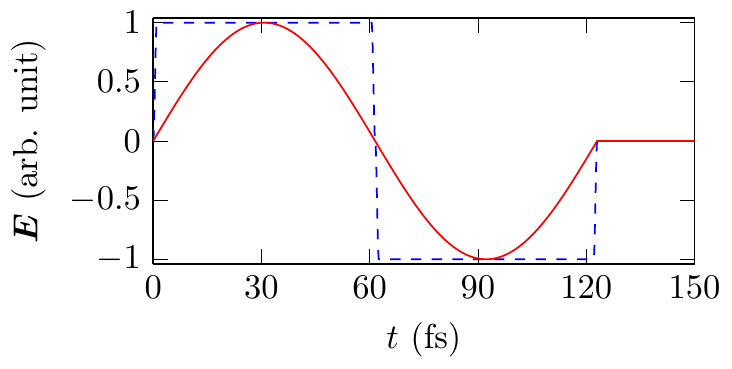}
	\caption{(Color online) Typical profiles of the electric fields used in the calculations. The blue dashed curve is referred as dc flash and the red solid curve as ac sine pulse. }
	\label{fig:field}
\end{figure}

Two types of generic fields for achieving control based on Stark shift mechanism are displayed in Fig.~\ref{fig:field}: A static dc field with positive or negative amplitudes, and a corresponding single optical cycle laser pulse with the same period satisfying Maxwell's equations requirement of a zero time-integrated area~\cite{Atabek}. The half period is taken as $60$ fs, basically in relation with typical decaying behaviors of the correlation function and the volume which display monotonic decrease at times later that $60$~fs with almost negligible values. The positive or negative amplitudes of the dc field are used to produce sudden negative or positive Stark shifts. As for the sine function describing the laser pulse, it is expected to provide an adiabatic excitation producing progressively the Stark shift which is looked for.

\begin{figure}[ht]
	\centering
	\includegraphics[width = 0.6\columnwidth]{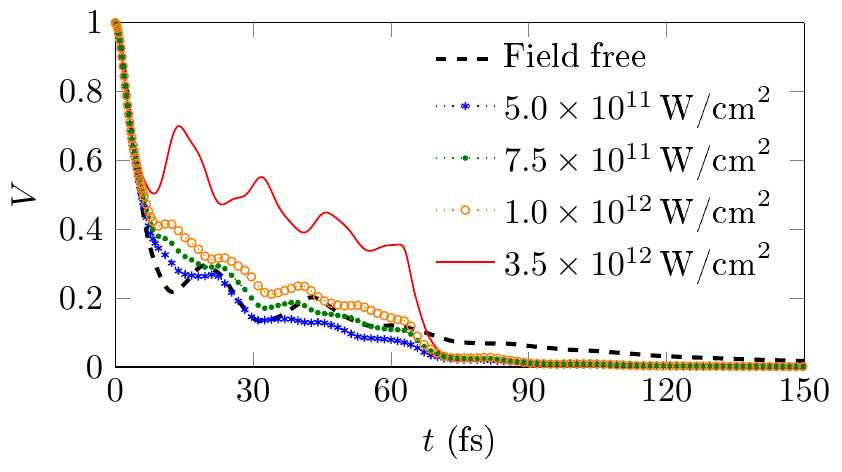}
	\caption{(Color online) Time evolution of the volume $ V $ for different intensities of ultra short dc flash.}
	\label{fig:vol_dc_flash}
\end{figure}

Inspired from and in conformity with the Fano-model analogy, we provide a numerical proof of the Stark shift mechanism on the non-Markovian control of the bath response by calculating the volume of accessible states for a collection of increasing dc fields intensities ranging from $5 \times 10^{11}$~W/cm$^2$ up to $3.5 \times 10^{12}$~W/cm$^2$. Such fields produce transition frequency gaps in the two-level system ranging from $0.015 au$ to $0.042 au$. When comparing with the field-free transition frequency $\omega_0 = 0.024 au$, we observe that both lower and higher shifts are observed (i.e., positive or negative contributions of the control fields). But most importantly, all situations which are depicted are non-resonant with respect to $\omega_{max} = 0.007 au$. It has to be emphasized that the Fano-type model interpretation can favor either positive or negative Stark shifts mediating off-the-energy-shell processes that are in competition in such non-resonant cases, putting the system transition frequency further or closer to the maximum spectral density frequency. The results of the corresponding time evolution of the volume are gathered in Fig.~\ref{fig:vol_dc_flash}. Several points are in order:
(i) For times $t \leq 60$~fs a noticeable enhancement of non-Markovianity is observed for intensities of the order of $10^{12}$~W/cm$^2$, both with respect to the slowing down of the overall decaying behavior and especially for the increasing amplitude of the bumps;
(ii) Such signatures are very much enhanced with stronger fields resulting into bumps with spectacular amplitudes reaching about $20\%$ of the initial value, and an overall decay which does not exceed $40\%$ at time $t=60$~fs, as compared with a value of about $10\%$ for the field-free case. The bump periodicity (about $15$~fs) is less than the one of the field-free case (about $20$~fs). This could be related with two effects; namely, the variation of the Rabi transition period of the central system, and, the indirect action of the control field on the bath;
(iii) For times exceeding $60$~fs, when the Stark shift becomes negative, still in conformity with Fano-model analogy, all previous dynamical behaviors are reversed, with a rather sudden decay much faster than the field-free case.
In summary, an efficient control of non-Markovianity referring to strong dc fields seems thus achievable, However, the experimental feasibility of such intense electric fields is questionable, despite their ultra short duration ($60$~fs) which makes them dc flashes rather than static dc fields. This is why we are now addressing Mid infrared or THz laser fields with the same periodicity (8 THz, 36$\mu$m wavelength) and comparable intensities, as illustrated in Fig.~\ref{fig:field} with the expectation that they will also provide efficient enough control tools.

\subsection{Laser control of non-Markovianity.}

\begin{figure}[ht]
	\centering
		\includegraphics[width = 0.6\columnwidth]{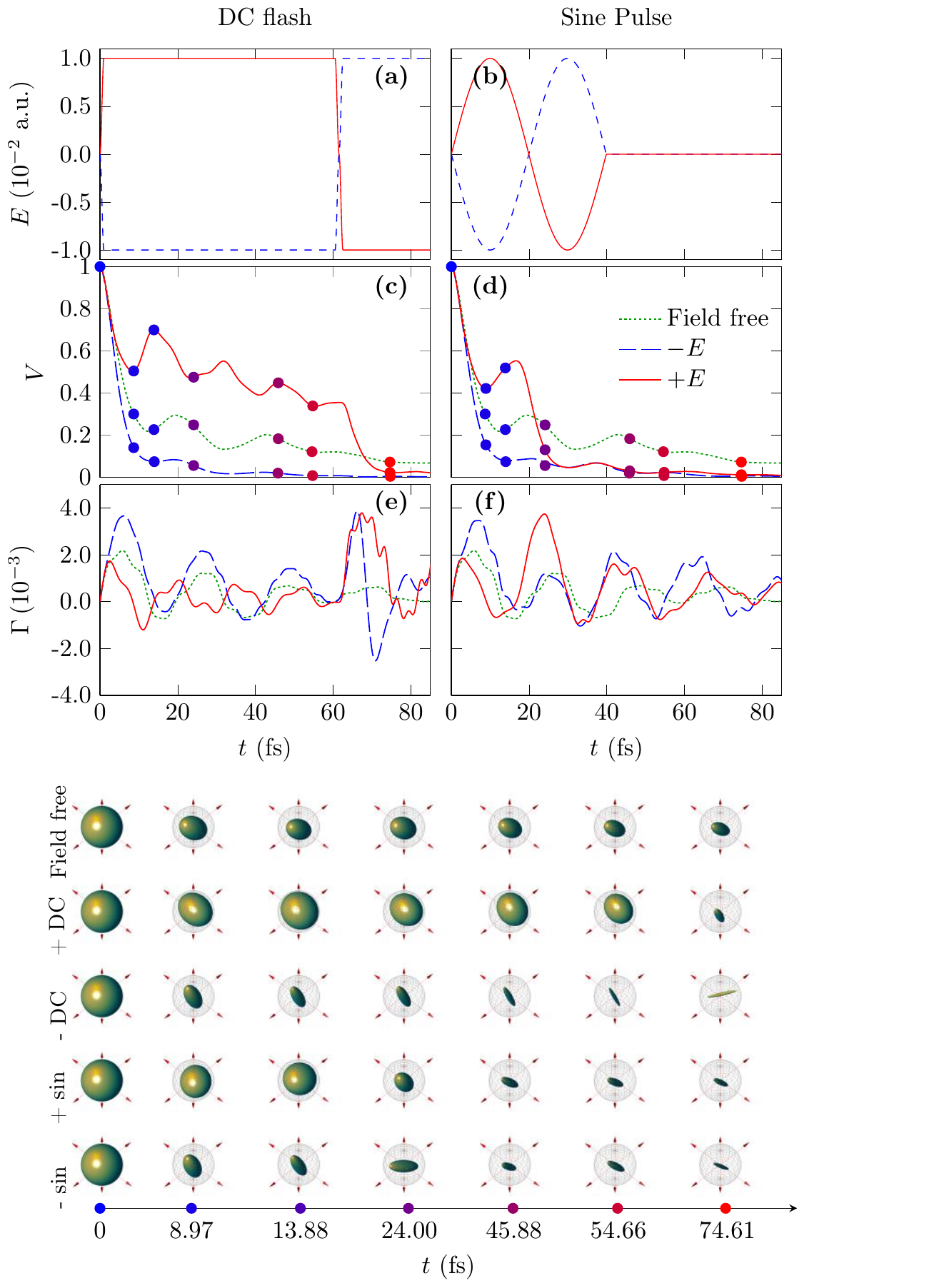}
		\caption{(Color online) Panels (a) and (b): Time evolution of the electric field amplitudes for a peak intensity of $ 3.5\times 10^{12} $ W/cm$ ^2 $. Panels (c) and (d): Time evolution of the volume of accessible states. Panels (e) and (f): Time evolution of the sum of canonical rates. The green dotted curve is for the field free case. The red and blue solid curves are respectively for the positive and negative initial values of the field amplitude. The lowest panel displays 3D illustrations of the Bloch ball  evolution (volume of accessible states) at specific times corresponding to the dots of panels (c) and (d). }
		\label{fig:vol_rate_field_direction}
\end{figure}

Typical control parameters being already fixed for ultra short dc flashes, we are now proceeding to more realistic laser control by referring to single optical cycle THz laser pulses of $12\mu$ wavelength and intensity of $3.5 \times 10^{12} W/cm^2$. Two phases are considered, leading to electric field amplitudes starting either with positive or negative values. Figure (\ref{fig:vol_rate_field_direction}) gathers the results for HEOM converged time evolution of the volume $V(t)$ [Eq.(\ref{volume})] and the total decoherence rate $\Gamma(t)$ [Eq.(\ref{gamma})]. We again observe the markedly different behaviors induced by positive or negative amplitudes rationalized by the off-resonant Fano-model analogy. In the present case, positive amplitudes (inducing negative Stark shifts) produce reduced transition frequencies increasing non-Markovianity, whereas negative ones (inducing positive Stark shifts) lead to very fast and more monotonous memory decay processes. But more importantly, THz laser pulses inducing a dynamical Stark shift of comparable amplitude with the ultra short dc flash, at least close to its maximum time, is actually shown to provide efficient non-Markovianity control. This is clearly proved by the spectacular enhancement of the first bump in the volume at about $20fs$. As compared with the field-free case the volume has almost gained a factor $\times 2$ of enhancement which could be considered as a very promising control achievement, solely and simply based on a comprehensive mechanism. In the lower panel of Fig.(\ref{fig:vol_rate_field_direction}), it is interesting to notice that the time-dependent behavior of the total decoherence rate $\Gamma(t)$ is quite close for the two dc or ac fields during the increasing amplitude period of the sine pulse, with temporary negative values around $15fs$ responsible for the most important bump in the volume. Clearly visible differences appear however for the decreasing amplitude part of the sine pulse.

Fig.~\ref{fig:vol_rate_field_direction} also displays the time evolution of the Bloch sphere representation of the volume of accessible sates, with the following mapping of the density matrix $\rho$ on the position vector $\vec{r}(x,y,z)$:
\begin{equation}
x=2\Re(\rho_{12}),
y=2\Im(\rho_{12}),
z=\rho_{22}-\rho_{11}.
\label{bloch}
\end{equation}
These are given at some specific times and help showing the different axes along which the volume decreases or temporarily increases building up the bumps we are referring to as signatures of non-Markovianity.
A complete dynamics of the ellipsoidal volume evolving inside the Bloch sphere is illustrated in Supplemental Material in terms of animated figures (see Supplemental Figs. S2 and S3).

In addition, we also try to get a better understanding of the control field dependence of the first moment of the collective mode in each electronic state given by the diagonal elements of the $X_1(t)$ matrix (Eq.~\eqref{X1}), emphasized as signatures of the field induced correlated system-bath dynamics.
\begin{figure}[ht]
	\centering
	\includegraphics[width = 0.6\columnwidth]{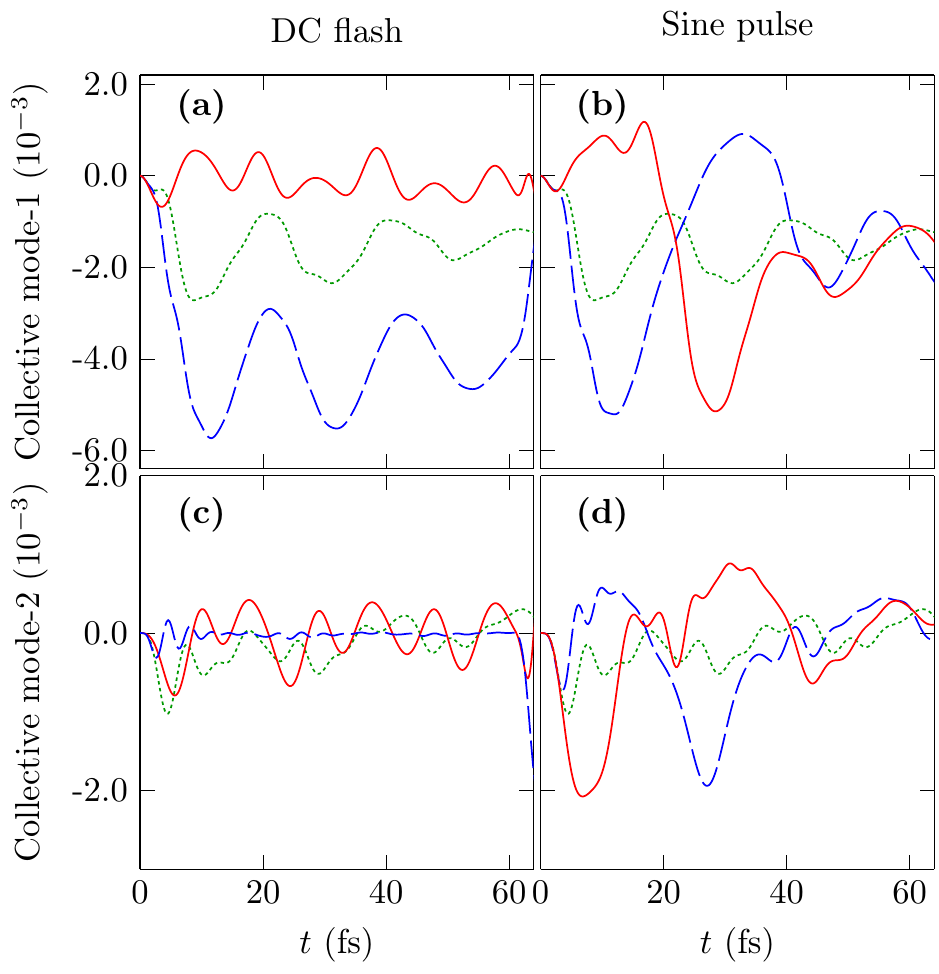}
	\caption{(Color online) Time evolution of first moments of the collective mode in each electronic state for an applied field of intensity $ 3.5\times 10^{12} $ W/cm$ ^2 $. The green dotted curve is the field free case, the red solid curve for the positive and the blue dashed curve for the negative values of the field amplitude.}
	\label{fig:moments_field_direction}
\end{figure}
Figure~\ref{fig:moments_field_direction} displays the results for the collective mode in each state (1 and 2)  as a function of time using the ultra short dc flashes and the laser sine pulses of the upper panel of Fig.~\ref{fig:vol_rate_field_direction}, together with their field-free behaviors. Expectation values of the collective mode in states 1 and 2  are enhanced by factors exceeding $\times 2$ close to laser pulse maximum ($t=10$~fs) with respectively negative (for mode 1) or positive (for mode 2) field strengths. Ultra short dc flashes produce similar effects on their full duration ($60$~fs). This shows in particular, how the strong laser interaction can modify the central system-bath couplings efficiently building some collective modes in the bath.

\subsection{Consequence of the control on the system characteristic observables.}

\begin{figure}[ht]
	\centering
	\includegraphics[width = 0.5\columnwidth]{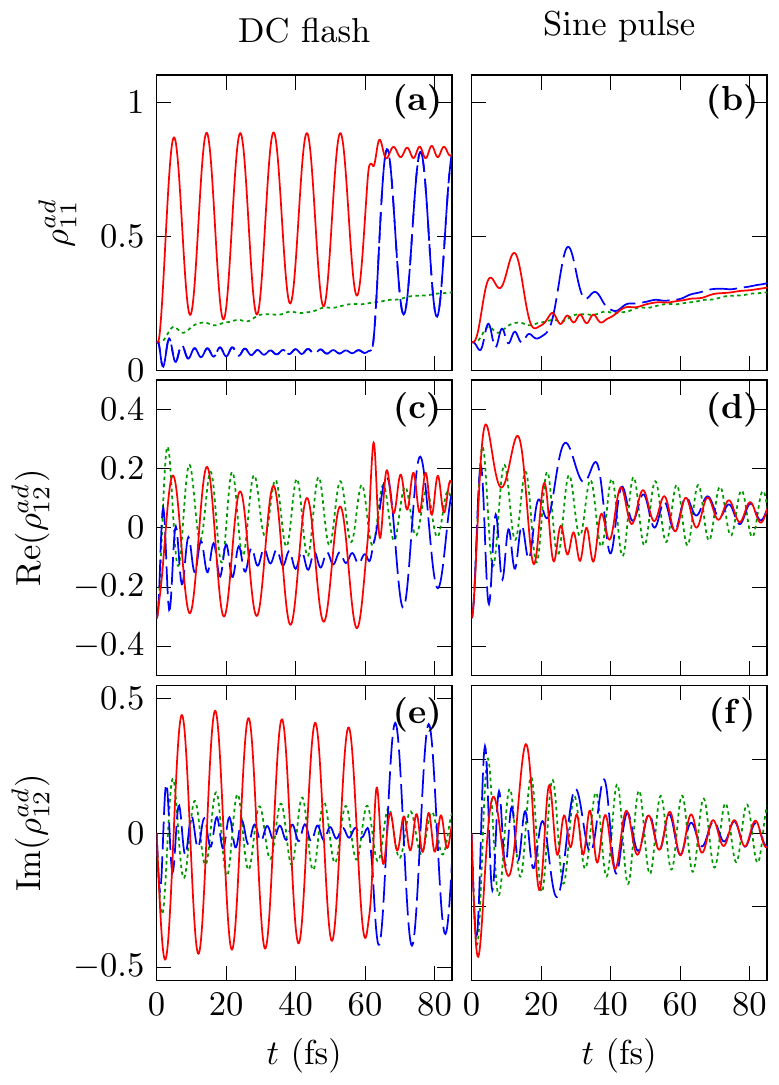}
	\caption{(Color online) Panels (a) and  (b) depict the adiabatic ground state population; (c) and (d) illustrate $ \Re(\rho^{ad}_{12}) $; (e) and (f) show $ \Im(\rho^{ad}_{12}) $ for the heterojunction excited with dc (left column) or ac (right column) fields. The green dotted curve is the field free case; the red solid one for the positive and the blue dashed one for the negative values of the field amplitude. All parameters are the same as in Fig.~\ref{fig:moments_field_direction}}
	\label{fig:coh_pop_dc_direction}
\end{figure}

Referring to the Stark shift in the central system transition frequency as a basic mechanism and controlling it through the intensity of a THz laser pulse, we have proceeded to an efficient control of the non-Markovian response of the bath. We wish now to analyze the consequences of such a control on the system, initial state dependent, physical observables. The initial population in the diabatic representation is ($\rho_{11}=1,~\rho_{22}=0$). The time-evolution of the population in the ground eigenstate (adiabatic state), together with the off-diagonal element of the density matrix in this representation, $\rho^{ad}_{12}$, as a signature of coherence, are displayed in Fig.~\ref{fig:coh_pop_dc_direction}. Positive dc fields rapidly increase $\rho^{ad}_{11}$ and induce typical population oscillations around $1/2$. But more interestingly, the amplitude of oscillations in the coherence terms $\rho^{ad}_{12}$ (both real and imaginary parts) are much increased.
Similar observations are valid for laser sine pulses, at least for times up to $20fs$ (i.e. half-cycle period), even though being much moderate. Such slowing down of decoherence can be connected with the bumps of the volume evolution and its ellipsoidal shape along ($x,y$)- axes of Fig.~\ref{fig:vol_rate_field_direction}.

\begin{figure}[ht]
	\centering
	\includegraphics[width = 0.6\columnwidth]{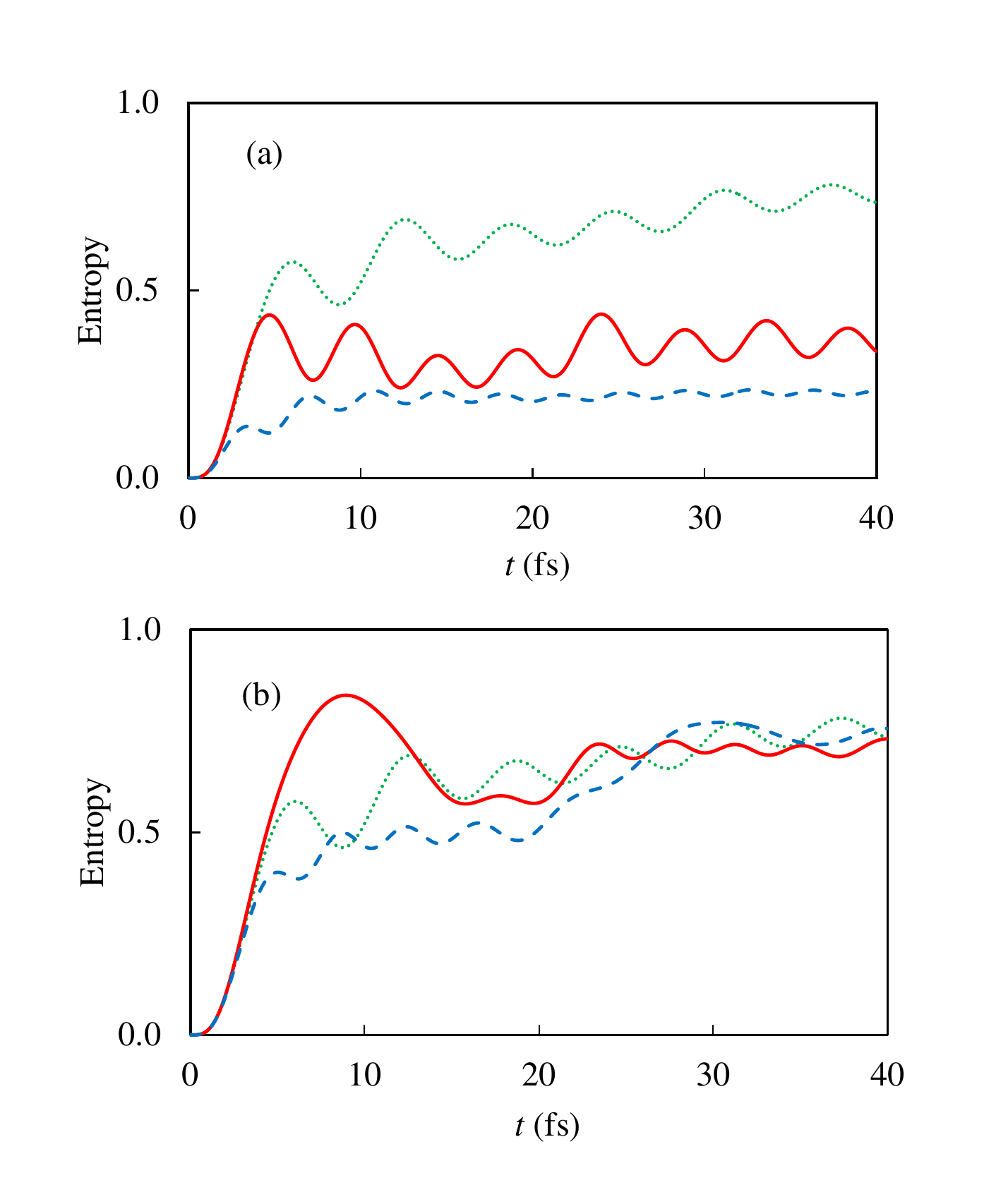}
	\caption{(Color online) Entropy of the system: Panel (a) for a 40fs ultra short duration dc flash and (b) for a sine pulse of 40fs period. Green dotted curve is for the field-free case, red solid curve is for the positive and blue dashed curve for the negative values of the field amplitude.}
	\label{fig:entropy}
\end{figure}

A last observable is the time-evolution of von Neumann's entropy of the central system given by \cite{Haseli}:
\begin{equation}
S(t)=-Tr[\rho(t)log_2\rho(t)]
\label{entropy}
\end{equation}
which is displayed in Fig.~\ref{fig:entropy}.
For ultra short dc flashes (both negative and positive amplitudes) as expected, the entropy is less than the one obtained without the control field all along the dynamics. The differences, temporarily more than a factor $\times 2$, could be qualified as spectacular. 
For sine pulses shaped following the dc ones (of period $40fs$ as in Fig.~\ref{fig:vol_rate_field_direction}) an entropy decrease, although much moderate, is still temporarily observed, for times around ($t=17fs$ to $23fs$). It is to be noted that these times precisely correspond to bump occurrence in the evolution of the volume of accessible states (Fig.~\ref{fig:vol_rate_field_direction}d) as a signature of non-Markovianity. Later on, apart from low amplitude oscillations, the evolution of the entropy is no more affected by the control field.
More unexpected is the short time ($t\leq 10fs$) evolution of the field controlled entropy showing an important increase and leading to values even higher than the ones of the field-free case. To rationalize such a behavior two points could be emphasized: (i) This short time dynamics is to be related with a fast decay of the volume, where the control field is not efficiently acting on the coherence (as is shown in the Supplemental Material \cite{SM}); (ii) More important is the fast decay of entropy between $t=10fs$ and $t=17fs$ (the negative slope of the red curve in Fig.~\ref{fig:entropy}) which turns out to be the observable that actually has to be considered as a consequence of non-Markovianity increase.

\section{Conclusion.}

In summary, we are aiming at developing theoretical tools for the laser control of OQS described by a spin-boson Hamiltonian and solving the Nakajima-Zwanzig master equation referring to a numerical method, involving HEOM at converged level of the hierarchy.
Two steps are followed for this goal: (i) address as a measure of non-Markovianity the non-monotonous decay of the volume of accessible states and, in particular, the occurrence and amplitude of bumps in its time evolution (i.e., negative total decay rates at particular times) \cite{lorenzo2013}, (ii) identify, in a comprehensive way, a basic laser induced mechanism, to enhance the non-Markovian bath response.
Such a response being generically obtained  through an appropriate laser pulse, we proceed to the time evolution of the system physical observables and in particular their coherence characteristics. This is done having in mind the following question: How much enhancing non-Markovianity slows down the decoherence in the density matrix decay, or the dynamical entropy increase?

The illustrative system is the well-documented heterojunction between fullerene and oligothiophene molecules at a fixed interfragment distance. All parameters entering the model, at the exception of the transition dipole taken in a reasonable typical range of magnitude, are the ones provided by previous works. The basic mechanism we are looking for is inferred from a Fano-type model analogy of two discrete levels facing a quasi-continuum with a structured density of levels. This qualitatively gives rise to some comprehensive view of the non-Markovianity increase, at least when the interstate transition frequency is off-resonant with respect to the spectral density function peak frequency. A static dc-field of appropriate intensity can obviously produce the Stark shift which is looked for \cite{molphys}. This being taken as a basic control mechanism, we accordingly shape a THz laser pulse dynamically providing comparable shifts and therefore non-Markovian enhancement.

As the most remarkable result of this paper, in a robust and experimentally achievable way, we are increasing by more than a factor $\times 2$ the amplitude of the bumps during the short time evolution of the volume of accessible states, enhancing thus considerably non-Markovianity. The analysis of the consequences on the two-level subsystem reduced density matrix and entropy or bath collective modes, shows, although at a rather modest level, a slowing down of decoherence signatures which are promising for future attempts concerned by optimal control \cite{Glaser,NJPoct2017}.
Actually this is the aim of our future project, that is referring to control theory for shaping a laser pulse such as to optimally slow down the decoherence dynamics of the central system populations (protection or reborn of the initial state, or efficient interstate switching) and/or the increase of its dynamical entropy, observe the consequences on non-Markovianity signatures, to show how the two control issues are closely related. We are actively pursuing our research in this direction.


\section*{Acknowledgment}
We acknowledge support from the ANR-DFG, under Grant No. \mbox{ANR-15-CE30-0023-01.} This work has been performed with the support of Technische Universit\"at M\"unchen Institute for Advanced Study, funded by the German Excellence Initiative and the European Union Seventh Framework Program under Grant Agreement No. 291763.


\bibliographystyle{apsrev4-1}

\begin{thebibliography}{10}
\bibitem{Weiss} Weiss U 2012 \textit{Quantum dissipative systems},
	World Scientific,4th Ed. Singapore
	
\bibitem{Breuer}  Breuer H-P. and Petruccione F 2002 \textit{The Theory of Open Quantum System},
	Oxford University Press

\bibitem{BreuerLPV} Breuer H-P, Laine E M, Piilo J and Vacchini B 2016 \textit{Rev. Mod. Phys.} \textbf{88} 021003
	
\bibitem{Rivas} Rivas A, Huelga S F and Plenio M B 2014 \textit{Rep. Prog. Phys.} \textbf{77} 094001

\bibitem{Vega} de Vega I and Alonso D 2017 \textit{Rev. Mod. Phys.} \textbf{89} 015001

\bibitem{Krauss} Krauss K 1983 \textit{States, Effects and Operations : Fundamental Notions of Quantum Theory}, Sringer Berlin
	
\bibitem{Alicki} Krauss K 1983 \textit{Quantum Dynamical Semigroups and Applications}, Sringer Berlin
	
\bibitem{MayKuhn} May V and K\"uhn O 2011 \textit{Charge and Energy Transfer in Molecular System} Wiley-VCH  Berlin
	
\bibitem{Glaser} Glaser S J, Boscain U, Calarco T, Koch C P, K\"ockenberger W, Kosloff R, Kuprov I, Luy B, Schirmer S, Schulte-Herbr\"uggen T, Sugny D, and F.K. Wilhelm F K 2015 \textit{Eur. Phys. J. D} \textbf{69} 279
	
\bibitem{koch2}  Koch C P 2016 \textit{J. Phys. Condens. Matter} C \textbf{28} 213001
\bibitem{Chin2013} Chin, A. W. and Prior, J. and Rosenbach, R. and Caycedo-Soler, F. and Huelga, S. F. and Plenio, M. B. 2013 \textit{Nature Phys.} \textbf{2}, 113
	
\bibitem{Gelinas2014} G{\'e}linas, S. and Rao, A. and Kumar, A. and Smith, S. L. and Chin, A. W. and Clark, J. and van der Poll, T. S. and Bazan, G. C. and Friend, R. H. 2014 \textit{Science} \textbf{343}, 512

\bibitem{Scholes2011} Scholes, G. D. and Fleming, G. R. and Olaya-Castro, A. and van Grondelle, R. 2011 \textit{Nat. Chem.} \textbf{3}, 763

\bibitem{Potz} P\"otz W 2006 \textit{Appl. Phys. Lett.} \textbf{89} 254102
	
\bibitem{Potz2} Wenin M and P\"otz W 2006 \textit{Appl. Phys. Lett.} \textbf{929} 103509
	
\bibitem{Lidar} Grace M, Brif C, Rabitz H, Walmsley I A, Kosut R I and Lidar D A 2007 \textit{J. Phys.} B \textit{At. Mol. Opt. Phys.} \textbf{40} S103
	
\bibitem{Cui} Cui W, Xi Z R and Pan Y 2008 \textit{Phys. Rev.} A  \textbf{77} 032117
	
\bibitem{Tai} Tai J-S, Lin K-T and Goan H-SE 2014 \textit{Phys. Rev.} A  \textbf{89} 062310



\bibitem{Koch} Reich D M, Katz N and Koch C P 2015 \textit{Sci. Rep.} 5:12430 DOI:10.1038/srep12430

\bibitem{Poggi} Poggi P M, Lombardo M C and Wisiacki D A 2017 \textit{EPL} \textbf{118} 20005

\bibitem{molphys} Puthumpally-Joseph R, Atabek O, Mangaud E, Desouter-Lecomte M, Sugny D 2017 \textit{Mol. Phys.} \textbf{115}, 1944
	
\bibitem{NJPoct2017} Mangaud E, Puthumpally-Joseph R, Sugny D, Meier C, Atabek O and Desouter-Lecomte M, \emph{Non-Markovianity in the optimal control of an open quantum system strongly coupled to a bath}, submitted to New J. Phys. (2017)

\bibitem{Meier}Meier C and Tannor D J 1999 \textit{J. Chem. Phys.} \textbf{111} 3365

\bibitem{Ohtsuki} Ohtsuki Y 2003 \textit{J. Chem. Phys.}  \textbf{119}  661

\bibitem{Yan} Xu R, Yan Y, Ohtsuki Y, Fujimura Y and Rabitz H 2004 \textit{J. Chem. Phys.}  \textbf{120} 6600

\bibitem{Leggett} Leggett A J, Chakravarty S,  Dorsey A T, Fisher M P A, Garg A and Zwerger 1987 W \textit{Rev. Mod. Phys.} \textbf{59} 1
	
\bibitem{Tamura}  Tamura H, Burghardt I and Tsukada M 2011 \textit{J. Phys. Chem.} C \textbf{115} 10205
	
\bibitem{Tamura2} Tamura H, Martinazzo R,  Ruckenbauer M and Burghardt I \textit{J. Chem. Phys.} 2012 \textbf{137}  22A540
	
\bibitem{Mangaud} Mangaud E, Meier C and Desouter-Lecomte M \textit{Chem. Phys.} 2017 \textbf{494} 90
	
\bibitem{Kubo} Tanimura Y and Kubo R 1989 \textit{J. Phys. Soc. Jpn.} \textbf{58} 101
	
\bibitem{Ishizaki} Ishizaki A and Tanimura Y 2005 \textit{J. Phys. Soc. Jpn.} \textbf{74} 3131

\bibitem{Tanimura} Tanimura Y 2006 \textit{J. Phys. Soc. Jpn.} \textbf{75} 082001
	
\bibitem{lorenzo2013} Lorenzo S and Plastina F and Paternostro M 2013 \emph{Phys.  Rev.  A} 88,  020102

\bibitem{SM} See Supplemental Material at [URL will be inserted by publisher] to illustrate the time evolution of the volume of accessible states

\bibitem{seideman2003} Stapelfeldt H.and Seideman T, 2003 \emph{Rev. Mod. Phys.} \textbf{75}, 543

\bibitem{Chenel2} Chenel A, Mangaud E,Burghardt I, Meier C and Desouter-Lecomte M 2014 \textit{J. Chem. Phys.}  \textbf{140} 044104

\bibitem{Pomyalov} Pomyalov A, Meier C and Tannor D J 2010 \textit{Chem. Phys.} \textbf{370} 98

\bibitem{Shi} Zhu L, Liu H, Xie W and Shi Q 2012 \textit{J. Chem. Phys.} \textbf{137} 194106

\bibitem{Fano} U. Fano, 1961 \emph{Phys. Rev.} \textbf{124}, 1866

\bibitem{Finkelstein} Finkelstein-Shapiro D, Urdaneta I,  Calatayud M, Atabek O, Mujica V and Keller A, 2015 \emph{Phys. Rev. Lett.} \textbf{115}, 113006

\bibitem{BreuerLP} Breuer H-P, Laine E M, Piilo J 2009 \textit{Phys. Rev. Lett.} \textbf{103} 210401

\bibitem{clos2012} Clos G and Breuer H-P, 2012 \emph{Phys. Rev. A} \textbf{86}, 012115

\bibitem{Paternostro} Lorenzo S, Plastina F and Paternostro M 2011 \textit{Phys. Rev.} A \textbf{84} 032124
	
\bibitem{Anderson} Hall M J W, Cresser J D,  Li L and Anderson E 2014 \textit{Phys. Rev.} A  \textbf{89} 042120
	
\bibitem{Anderson2} Anderson E, Cresser J D and Hall M J W  2007 \textit{J. Mod. Phys.} \textbf{54} 1695
	
\bibitem{Haseli} Haseli S, Salimi S and Khorashad A S 2015 \textit{Quant. Inf. Proc.} \textbf{14} 3581

\bibitem{Atabek} Sugny D, Vranckx S, Ndong M, Vaeck N, Atabek O and Desouter-Lecomte M 2014 \textit{Phys. Rev.} A  \textbf{90} 053404




	

	


	
\end{thebibliography}

\begin{thebibliography}{11}
	\bibitem{S_lorenzo2013} Lorenzo S and Plastina F and Paternostro M 2013 \emph{Phys.  Rev.  A} 88,  020102
\end{thebibliography}


\pagebreak
\widetext
\begin{center}
	\textbf{\large Supplemental Materials: Basic mechanisms in the laser control of non-Markovian dynamics.}
\end{center}
\setcounter{equation}{0}
\setcounter{figure}{0}
\setcounter{table}{0}
\setcounter{section}{0}
\setcounter{page}{1}
\makeatletter
\renewcommand{\theequation}{S\arabic{equation}}
\renewcommand{\thefigure}{S\arabic{figure}}
\renewcommand{\bibnumfmt}[1]{[S#1]}
\renewcommand{\citenumfont}[1]{S#1}
\vspace{1cm}

A time dependent animation of the Bloch sphere representation of the density matrix evolution gives a clear view of the way the control process is operating.

\noindent
The density matrix $\rho$ is mapped as a time dependent position vector $\vec{r}$ with coordinates $\vec{r}(x,y,z;t)$ such that \cite{S_lorenzo2013}:
\begin{equation}
	x=2 \Re{\rho_{12}}, \ \ \ \ y=2 \Im{\rho_{12}}, \ \ \ \ z=\rho_{22}-\rho_{11}
\end{equation}
where $x$ and $y$ describe coherence, whereas $z$ addresses populations.
We are hereafter providing two illustrations dealing with typical THz control fields as referred to in the main text and which we claim to produce generic results.

 \section{Bloch vector dynamics.}
 
 \begin{figure}[ht]
 	\centering
 	\includegraphics[angle=-90,width=\textwidth]{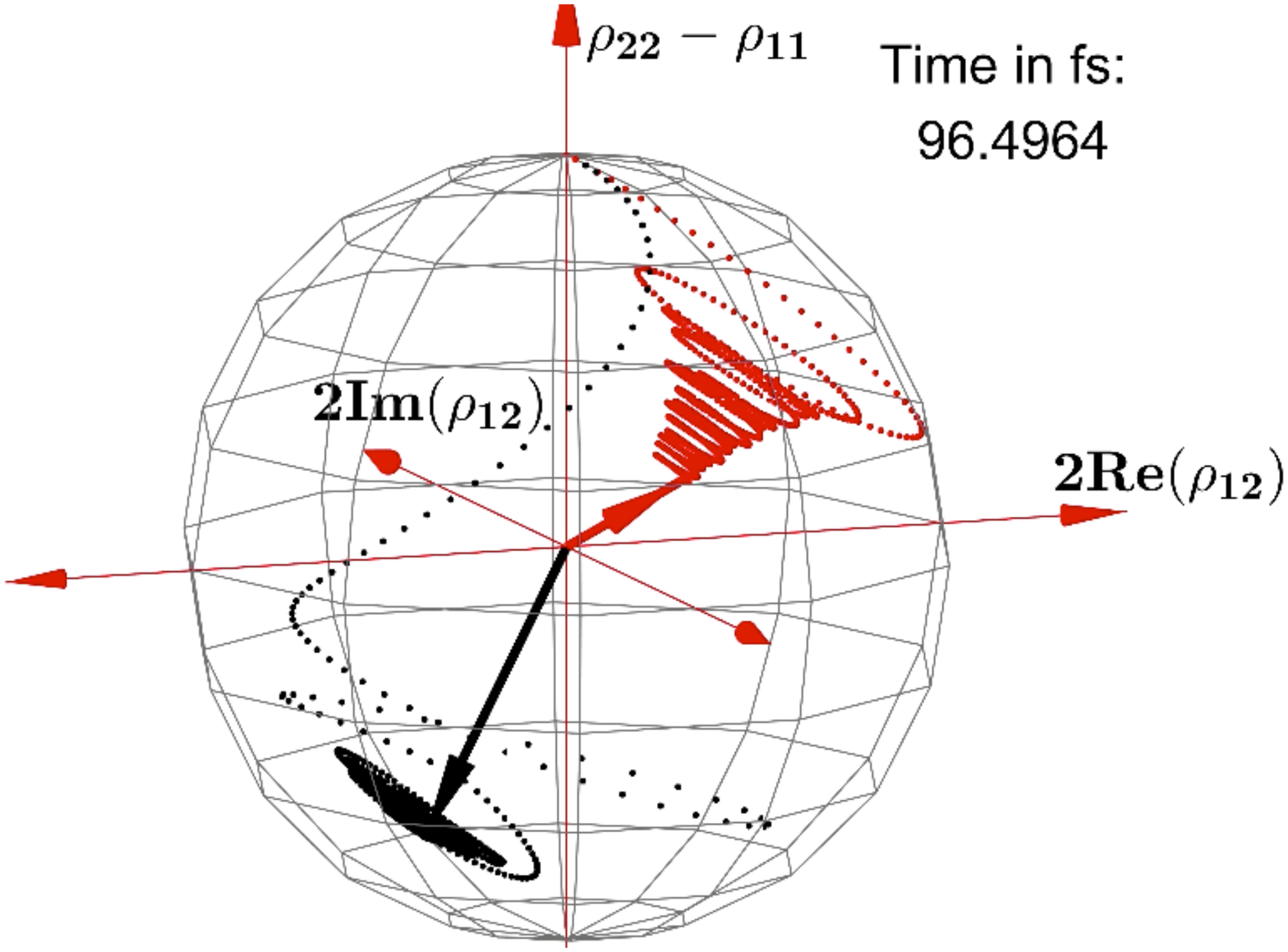}
 	\caption{(Color online) 3D representation of the trajectories of the Bloch vector up to 100fs. The red and black arrows are for the field-free and typical control fields respectively.}
 	\label{fig:trajectories}
 \end{figure}
\noindent 
 The first observable concerns the time dependent behavior of the Bloch vector $\vec{r}(t)$ when starting from the North pole of the unitary sphere (i.e., $z=1$, with $\rho_{22}=1$ and $\rho_{11}=0$) taken as an initial condition.
This is displayed in Figure \ref{fig:trajectories} up to $t=200 fs$ corresponding to typical complete decay times of the volume of accessible states. The Bloch vector is graphically represented by an arrow pointing in the corresponding polar angular direction, with its length proportional to its geometrical norm, decaying from unity (at initial time $t=0$) to some smaller values as time evolves, but not systematically in a monotonous way.
The arrowhead follows a trajectory starting from the North pole of the unitary Bloch sphere, and remains embedded in it later on. Two dynamics are depicted: In red for the field-free case and in black when attempting control through an external laser field with, as an objective, non-Markovianity enhancement measured by the increase of bump amplitudes in the time evolution of the volume of accessible states. 

\noindent
In the field-free case, short time dynamics shows a fast norm decrease, mainly along $z$-axis (i.e., population). This is followed by a vortex type behavior around an average position far from the North pole and corresponding to a vector whose projection in the ($x,y$)-plane is close to the bisector. Later, the evolution moves $\vec{r}(t)$ toward the eye of the vortex with a rather fast decay along ($x,y$)-axes, resulting into noticeable decoherence already at times about $t=50fs$.

\noindent
Completely different dynamics is observed when applying the control field. Starting again from the North pole, $\vec{r}(t)$ rapidly changes sign and is further evolving close to the South pole, but with much moderate norm decrease. Later dynamics is also characterized by a vortex behavior but now with an average position much closer to the South pole than the one of the field-free case was to the North pole. More importantly, the vortex widens considerably less, showing thus much better and longer coherence conservation, and as a consequence non-Markovianity.

 \section{Bloch ball dynamics.}
 
 \begin{figure}[ht]
 	\centering
 	\includegraphics[angle=0,width=\textwidth]{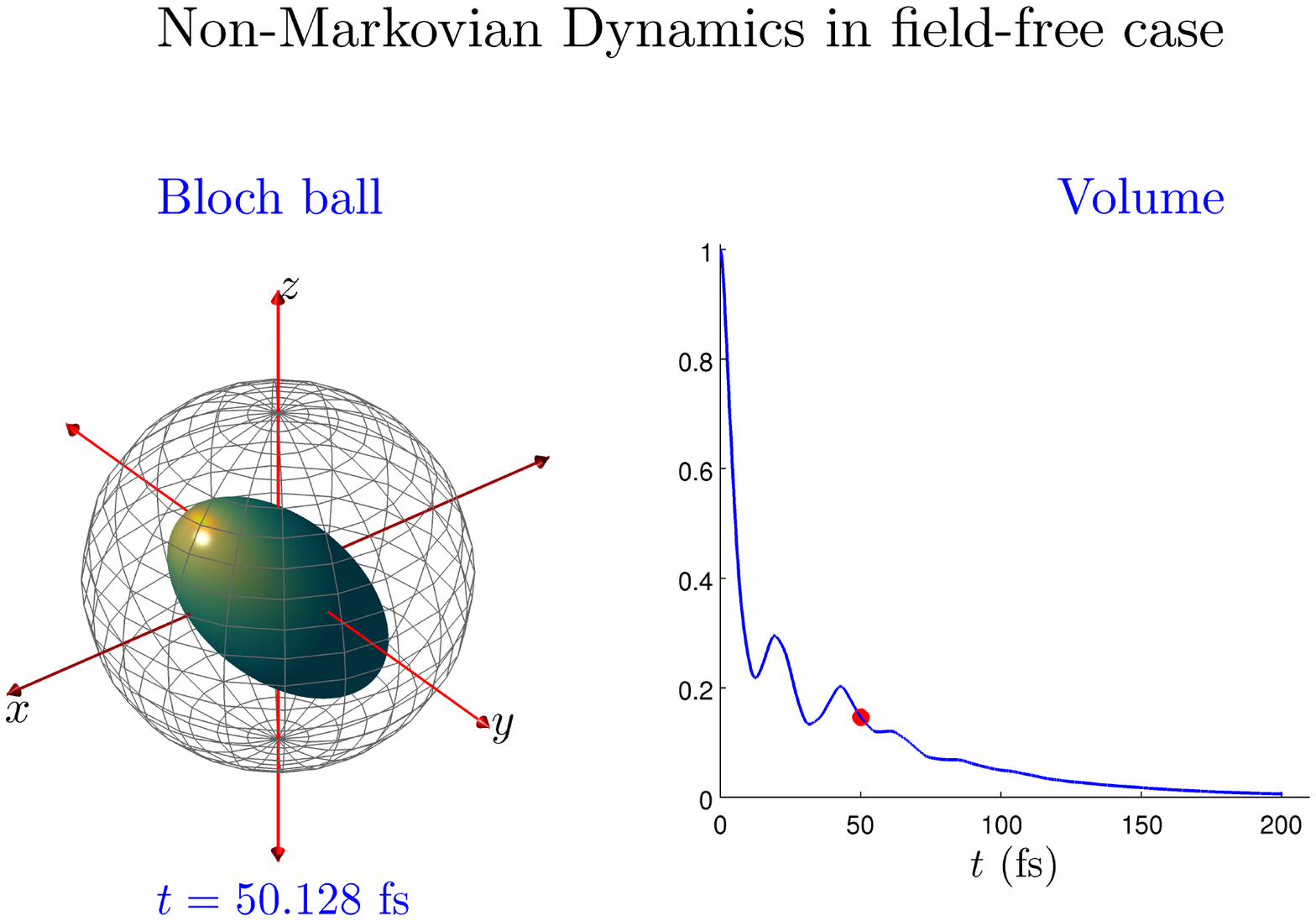}
 	\caption{(Color online) Right panel: 3D representation of the Bloch ball up to 200fs in the field-free case. Left panel: Time evolution of the volume of accessible states.}
 	\label{fig:shape_without_field}
 \end{figure}
 \noindent
 The second observable is the volume of all accessible states when starting from any initial condition on the unitary Bloch sphere. In other words, we are examining the 3D representation of the deformation of the Bloch ball leading to a volume, the surface of which is a mapping of the density matrix. The field-free case is shown in Figure \ref{fig:shape_without_field}, gathering the time evolution of the 3D ball with its scalar volume $V$. In our specific heterojunction inspired example $V$ is not monotonously decaying. Low amplitude bump structures are signatures of some non-Markovian dynamics. Our aim is to enhance them through an external laser field control. Short time dynamics shows a volume which is decreasing fast mainly along $z$-axis. This leads to an ellipsoidal ball pointing almost along $y$-axis at $t=6fs$. Later and before the occurrence of the bump ($t=14fs$) the ellipsoid main large axis is oriented along a central direction roughly corresponding to spherical angles ($\phi=\pi/2, \theta=\pi/2$). On the bump ($t=20fs$), the stretching along $z$-axis is still decreasing but those along $x$ and $y$-axes are now increasing. So, basically, the bump structure is in relation with a slowing down of decoherence (i.e;, increase of $\rho_{12}$). For long enough times, and before the volume collapses to zero, the ellipsoid is stretched along the $y$-axis.
 
 \noindent
When switching the control field on, what basically happens is that for short time dynamics the rather fast collapse along $z$-axis is much delayed. The volume remains close to a spheroidal ball. The bumps are much more pronounced, their amplitudes being about 4 times the ones of the field free case. Here again the ellipsoid is stretched along the $x$, $y$-axes ultimately showing that control enhancing non-Markovianty is mainly acting to protect the system against decoherence. 
	\begin{figure}[ht]
		\centering
		\includegraphics[angle=0,width=\textwidth]{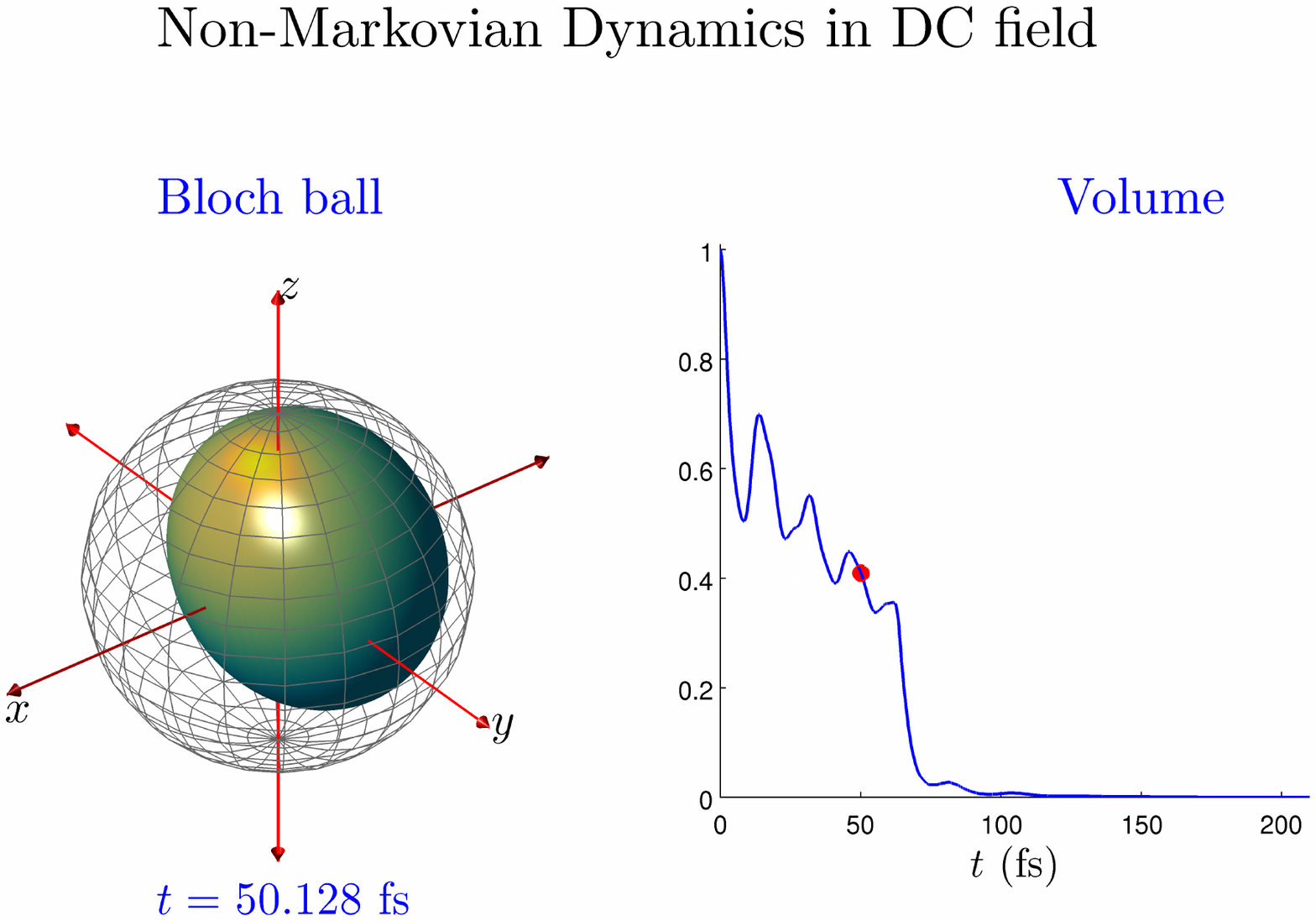}
		\caption{(Color online) Same as for figure \ref{fig:shape_without_field} but when acting with a typical external control field.}
		\label{fig:shape_with_field}
	\end{figure}
	
	\noindent
	Finally, the graphical lecture that we get from both the dynamics of the initial-state-dependent trajectories of the Bloch vector inside the unitary sphere and from the time evolving shape of the Bloch ball itself involving all initial states can be summarized as follows. The control fields shaped as to enhance non-Markovianity are acting at short times in such a way to delay the population decrease. As for later dynamics, it is dominated by a slowing down of the decoherence processes.

\end{document}